\documentclass[conference]{IEEEtran}
\usepackage{amsmath, amssymb, bm, cite, epsfig, psfrag}
\usepackage{graphicx}
\usepackage{soul} 
\usepackage[margin=0.7in]{geometry}
\usepackage{dblfloatfix}
\usepackage{array}
\usepackage{lipsum}
\newcolumntype{P}[1]{>{\centering\hspace{0pt}}p{#1}}
\newcolumntype{M}[1]{>{\centering\hspace{0pt}}m{#1}}
\newcolumntype{L}{>{\centering\arraybackslash}m{3cm}}
\usepackage[font=small]{caption}
\usepackage{epstopdf}	
\usepackage{longtable}
\usepackage{supertabular,booktabs}
\usepackage{bbm}
\usepackage{multirow}
\usepackage[usenames,dvipsnames]{xcolor}
\renewcommand{\arraystretch}{1.5}
\usepackage{subfigure}
\usepackage{etoolbox}
\usepackage{pbox}
\usepackage{fixltx2e}
\usepackage{tabu}	
\usepackage{filecontents}
\usepackage{enumerate}
\usepackage{textcomp}

\usepackage{colortbl}
\usepackage{fancyhdr}

\usepackage{bm}
\newtoggle{conference}
\togglefalse{conference} 
\graphicspath{{figures/}}

\fancypagestyle{firststyle}{
	\fancyhf{}
	\fancyhead[L]{Y. Xing and T. S. Rappaport, ``Propagation Measurement System and Approach at 140 GHz--Moving to 6G and Above 100 GHz,''  2018 \textit{IEEE Global Communications Conference (GLOBECOM)}, Abu Dhabi, UAE, Dec. 2018, pp. 1-6.}     

}

\begin{document}
\bibliographystyle{IEEEtran}
\title{Propagation Measurement System and Approach at 140 GHz--Moving to 6G and Above 100 GHz}
\author{\IEEEauthorblockN{Yunchou Xing, Theodore S. Rappaport}

\IEEEauthorblockA{	\small NYU WIRELESS\\
					NYU Tandon School of Engineering\\
					Brooklyn, NY 11201\\
					\{ychou, tsr\}@nyu.edu}}

\maketitle
\thispagestyle{firststyle}

\begin{abstract}
With the relatively recent realization that millimeter wave frequencies are viable for mobile communications, extensive measurements and research have been conducted on frequencies from 0.5 to 100 GHz, and several global wireless standard bodies have proposed channel models for frequencies below 100 GHz. Presently, little is known about the radio channel above 100 GHz where there are much wider unused bandwidth slots available. This paper summarizes wireless communication research and activities above 100 GHz, overviews the results of previously published propagation measurements at D-band (110-170 GHz), provides the design of a 140 GHz wideband channel sounder system, and proposes indoor wideband propagation measurements and penetration measurements for common materials at 140 GHz which were not previously investigated.
\end{abstract}
    
\begin{IEEEkeywords}
    mmWave; 5G; D-band; 6G; Channel sounder; 140 GHz propagation measurements; Terahertz (THz)
\end{IEEEkeywords}

\section{Introduction}~\label{sec:intro}
Fifth generation (5G) wireless communication systems will use millimeter wave (mmWave) frequency bands (30-300 GHz) to offer unprecedented spectrum and muti-Gigabit-per-second (Gbps) data rates to a mobile device \cite{rappaport2013millimeter, Rap02a}. Early work showed that 15 Gbps peak rates are possible with $4 \times 4$ antenna phased arrays at the mobile handset and 200 m spacing between the base stations (BSs) \cite{ghosh2014mmwave, roh2014millimeter}. The U.S. Federal Communications Commission (FCC) authorized its 2016 ``Spectrum Frontiers'' with unprecedented allocations of 10.85 GHz of mmWave spectrum for 5G advancements \cite{FCC16-89}, and in September 2017, the 3rd Generation Partnership Project (3GPP) proposed a technical specification document on new radio (NR) access technology to support 5G networks \cite{3GPPNR}. 

Extensive measurements and research have been conducted around the world in the 28 GHz, 38 GHz, 60 GHz and 73 GHz frequency bands for 5G cellular systems and 5G WiFi networks \cite{rappaport2013millimeter, maccartney2015indoor, 802.11ad}. Standard bodies and organizations such as 3GPP, 5G Channel Model (5GCM), Mobile and wireless communications Enablers for the Twenty-twenty Information Society (METIS), and Millimeter-Wave Based Mobile Radio Access Network for Fifth Generation Integrated Communications (mmMAGIC) have proposed channel models for frequencies below 100 GHz based on extensive field data collected over the past few years \cite{3GPP2017, 5GCM, METIS2015, mmMAGIC, Sun18a}. Various companies have conducted 5G field trials. For example, AT\&T has achieved 1.2 Gbps at a mobile user in a 400 MHz channel (28 GHz band) with 9-12 milliseconds (ms) latency at more than 150 meters away from the cell site, which is a huge improvement compared to the  AT\&T's current 4G average download speed of 15 Mbps with 58 ms latency \cite{ATTtrials}. Verizon has been trialing fixed 5G in eleven cities at 28 GHz and 39 GHz in the U.S., with fixed 5G service progressing better than expected. Due to these successful trials and testbeds, 2018 will likely be the first year of commercialization for 5G. 

However, little is known about radio channels above 100 GHz where there are wider unused bandwidth slots available. The immense bandwidths at mmWave and THz frequencies can enable future indoor and outdoor mobile networks as well as rural macrocell (RMa) point-to-point copper replacement over very large distances \cite{Mac16c, Mac17JSACa}. For example, there is 60 GHz of spectrum in D-band (110 GHz to 170 GHz) and when allocated for high-speed wireless links, this large bandwidth has potential applications in ``wireless fiber'' backhaul for fixed links, indoor/WiFi access, mobile communication, precision positioning, velocity sensors, passive mmWave cameras, vehicular communication, navigation, radar, and on-body communication for health monitoring systems \cite{Rap15a, kim2015Dband,ojas2018globecomm}.

\begin{figure}    
	\centering
	\includegraphics[width=0.5\textwidth]{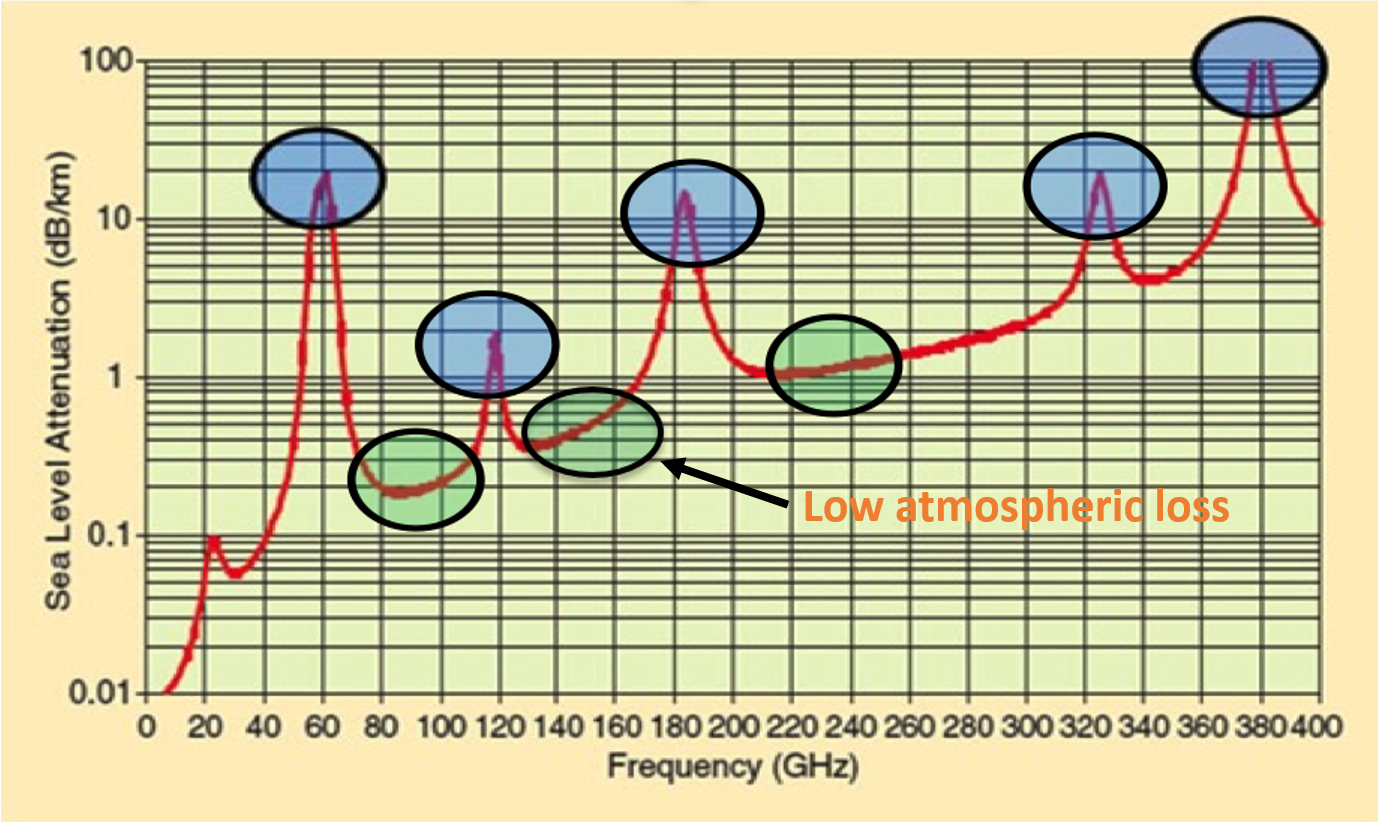}
		\vspace{-1.5 em}
	\caption{Atmospheric absorption of electromagnetic waves at sea level versus frequency, showing the additional path loss beyond free space propagation due to atmospheric absorption \cite{rappaport2011state}. }
		\vspace{-1.5 em}
	\label{fig:airabsorb}
\end{figure}

\section{Moving to 6G and frequencies above 100 GHz}

Various frequency bands suffer different atmospheric absorptions (e.g., oxygen and water molecule absorptions) resulting in additional path loss beyond free space propagation \cite{APS2, ITU-Rattenuation}. In addition to existing widespread agreement on spectrum allocation for short-range applications in the 60 GHz band, the 120, 183, 325, and 380 GHz bands (blue circles in Fig. \ref{fig:airabsorb}) are likely to be used for very close-in communications and ``whisper radio'' applications that replace wiring harnesses in circuit boards and vehicles, where massive bandwidth channels with zero error rate (due to coding, redundancy, and frequency diversity) will attenuate very rapidly beyond a few meters \cite{APS2}. Fig. \ref{fig:airabsorb} also shows the 77, 140, and 240 GHz bands (green circles), which only suffer 1 dB or less additional loss than caused by free space propagation per km in air \cite{rappaport2011state}, are suitable for longer-range broadband mobile and fixed applications.

 The increasing mmWave applications and technologies have stimulated interest and concerns about biological safety at mmWave frequencies. Work in \cite{wu2015safe,wu15ICC} summarized potential biological effects of nonionizing mmWave radiation on the human body and showed that even the eyes and skin, where these tissues would receive the most intense heat from radiation, would not suffer damage from mmWave in the far field.

The FCC initiated a proceeding (ET Docket No.18-21) to expand access to spectrum above 95 GHz for new services and technologies in February 2018 \cite{FCC18-21}. The notice of proposed rulemaking aims to seek comments on adopting rules for fixed point-to-point use of up to 102.2 GHz of licensed spectrum, making up to 15.2 GHz of spectrum available for unlicensed use, and creating a new category of experimental licenses for the 95 GHz to 3 THz range. In addition to FCC's rulemaking above 95 GHz, there are also other activities on the spectrum above 95 GHz. In 2014, Japan's Ministry of Internal Affairs and Communications (MIC) officially revised its regulations to allocate an 18 GHz wide band from 116 GHz to 134 GHz for broadcasting service, which is the first industrial allocation above 100 GHz carrier frequencies \cite{nagatsuma2014breakthroughs}. 

In 2015, The European Telecommunication Standards Institute (ETSI) Industry Specification Group (ISG) proposed a millimeter Wave Transmission (mWT) document on wireless transmission applications and use cases that can be addressed by mmWave spectrum, focusing on frequency bands from 50 GHz up to 300 GHz \cite{ETSIGSmWT}.

 The International Telecommunication Union (ITU) plans to identify frequency bands for global land-mobile and fixed services applications operating in the frequency range 275-450 GHz \cite{marcus2016WRC} in its 2019 World Radio communication Conference (WRC-19). Some research has already been conducted on frequencies above 100 GHz, and some of the recent results of high-speed mmWave and THz wireless communication systems are summarized in \cite{yu2016400,ma2017frequency,mumtaz17TVT}.

Optical fiber communication technologies have enabled spectrally efficient high data rate communications in wired networks, and 200 Gigabit Ethernet (GbE) and 400 GbE links will be deployed soon, supporting a speed of up to 400 Gbps \cite{802.3bs}. However, the installation and maintenance costs of optical fiber can be huge with an insufficient return on investment, especially in rural areas. In these cases, wireless point-to-point terrestrial communications with extremely high bandwidth will help to achieve rates comparable to the optical fiber while offering lower time latency than fiber \cite{Rap15a}. Work in \cite{Mac16c, Mac17JSACa} shows that surprisingly long distance (greater than 10 km) can be achieved in clear weather with less than 1 W of power at 73 GHz. A new rural macrocell (RMa) path loss model (CIH model), which is more accurate and easier to apply for varying transmitter antenna heights than the existing 3GPP/ITU-R RMa path loss models, is provided \cite{Mac17JSACa}. To connect rural America with upgraded access speeds, the FCC is now accepting applications from broadband providers of all kinds to participate in this summer's Connect America Fund Phase II reverse auction (AU Docket No. 17-182) \cite{FCC18-5} which will make available up to \$ 1.98 billion in support over the next decade to help build out high-speed Internet access to 1 million homes and small businesses in rural areas across the U.S. that lack service. 

Shrinking cell size has been proven to increase the spectrum efficiency and the total network capacity by reusing the spectrum \cite{chandrasekhar2008femtocell}. Nomadic base stations (BSs), direct device-to-device (D2D) connections, and massive Internet of things are envisioned to emerge in 5G for even greater capacity per user \cite{APS2}. The deployment of small cells, which are low-powered wireless BSs that cover smaller areas (e.g., homes, stadiums, and metropolitan outdoor spaces) than larger macrocells, will grow rapidly in number as 5G is deployed in 2018 and beyond. The FCC proposed a wireless infrastructure order (WT Docket No.17-79) \cite{FCC17-79} on March 2018 to accelerate wireless broadband deployment by removing barriers to small cell infrastructure deployment and investment. The order predicts a savings of \$ 1.56 billion between 2018-2026 which would create 57,000 small cells and to create 17,000 jobs beyond what would occur without the acceleration \cite{FCC17-79}.

\begin{table*}[!ht]\footnotesize
	\renewcommand{\arraystretch}{1.1}
	\centering
	\caption{140 GHz Band Channel Sounder} \label{tab:CS}
	\newcommand{\tabincell}[2]{\begin{tabular}{@{}#1@{}}#2\end{tabular}}
	\begin{tabular}{|c|c|c|c|c|}
		\hline
		\tabincell{c}{\textbf{ } }&\tabincell{c}{\textbf{Channel Sounder} \\ \textbf{Type}}&\tabincell{c}{\textbf{Bandwidth} }&\tabincell{c}{\textbf{Antennas} }&\tabincell{c}{\textbf{Dynamic range}}\\
		\hline
		\tabincell{c}{\textbf{Aalto University \cite{nguyen2017comparing,nguyen2016dual}}}&\tabincell{c}{Vector Network Analyzer}&\tabincell{c}{\textcolor{black}{Swept across} 4 GHz}&\tabincell{c}{19 dBi horn (RX) \\ and 2 dBi bicone (TX) antennas }&\tabincell{c}{130 dB \\ 3 - 65 m  \\ Indoor shopping mall environment }\\
		\hline
		\tabincell{c}{\textbf{ Georgia Institute} \\ \textbf{of Technology \cite{kim2015Dband,cheng2017comparison} }}&\tabincell{c}{Vector Network Analyzer}&\tabincell{c}{100 kHz}&\tabincell{c}{23 dBi horn antennas \\ (TX and RX) }&\tabincell{c}{90 dB \\ 0.3-1.8 m }\\
		\hline
		\tabincell{c}{\textbf{ NYU WIRELESS \cite{Mac17JSACb} }}&\tabincell{c}{Sliding correlation}&\tabincell{c}{4 GHz}&\tabincell{c}{27 dBi horn antennas\\ (TX and RX) }&\tabincell{c}{145 dB \\1-45 m \\ Indoor office NLOS environment }\\
		\hline
	\end{tabular}
\end{table*}

\section{Propagation in D-band (110-170 GHz)}

D-band is one of the most attractive frequency bands in the coming decade since there is 60 GHz of spectrum which can be used in applications requiring ultra-high bandwidth. \textcolor{black}{Relatively few public works exits about the propagation characteristics in D-band.} Propagation measurements in the 140 GHz band were conducted in a shopping mall \cite{nguyen2017comparing,nguyen2016dual} by Aalto University using a vector network analyzer (VNA) based channel sounder (LO multiplication factor is 12), where the transmitter (TX) and receiver (RX) were connected with a 200 m optical fiber cable. A 19 dBi horn antenna with a 10\textdegree~half power beam width (HPBW) in the azimuth plane and a 40\textdegree~HPBW in the elevation plane was used at the RX, and a bicone antenna, which was omnidirectional (2 dBi) in the azimuth plane and had a 60\textdegree~HPBW in the elevation plane, was used at the TX. With such antennas and -7 dBm input power, the channel sounder system had a 130 dB dynamic range (corresponding to a measurable link distance range of 3--65 m) at 140 GHz band with a 4 GHz bandwidth \cite{nguyen2017comparing}. 

A VNA sweeps discrete narrowband frequency tones across the bandwidth of interest to measure the S21 parameter of the wireless channel, followed by an inverse discrete Fourier transform of the channel transfer function, which results in a complex channel impulse response \cite{Mac17JSACb,Rap02a}. Due to its long sweep time across a broad spectrum which can exceed the channel coherence time, VNA based channel sounders are typically used in a static environment and require a cable that is a tripping hazard, which is likely why Aalto University conducted measurements at the shopping mall at night with an optical fiber cable connected.

Omnidirectional path losses at 140 GHz and 28 GHz were compared using the $\alpha-\beta$/floating-intercept (AB/FI) model, and work in \cite{nguyen2017comparing} showed that except for some additional free space path loss (FSPL) at 140 GHz, the slope (path loss exponent) and variations of the path loss data of the two bands were similar, which agreed with the findings in \cite{maccartney2015indoor} for 28 and 73 GHz using the close-in (CI) path loss model with 1 m reference distance. In addition, work in \cite{nguyen2017comparing} found that in the 140 GHz band, the numbers of clusters and multipath components (MPCs) in each cluster (an average of 5.9 clusters and 3.8 MPCs at 140 GHz) were fewer compared to the 28 GHz band (an average of 7.9 clusters and 5.4 MPCs) \cite{nguyen2017comparing}. While in 3GPP New Radio (NR) model TR 38.901 \cite{3GPP2017},  there were 15 clusters for line-of-sight (LOS) and 19 clusters for non-line-of-sight (NLOS) scenarios (with 20 MPCs per cluster), which were shown to be unrealistically high \cite{rappaport2017VTC, Sun18a}.

\begin{figure}    
	\centering
	\includegraphics[width=0.5\textwidth]{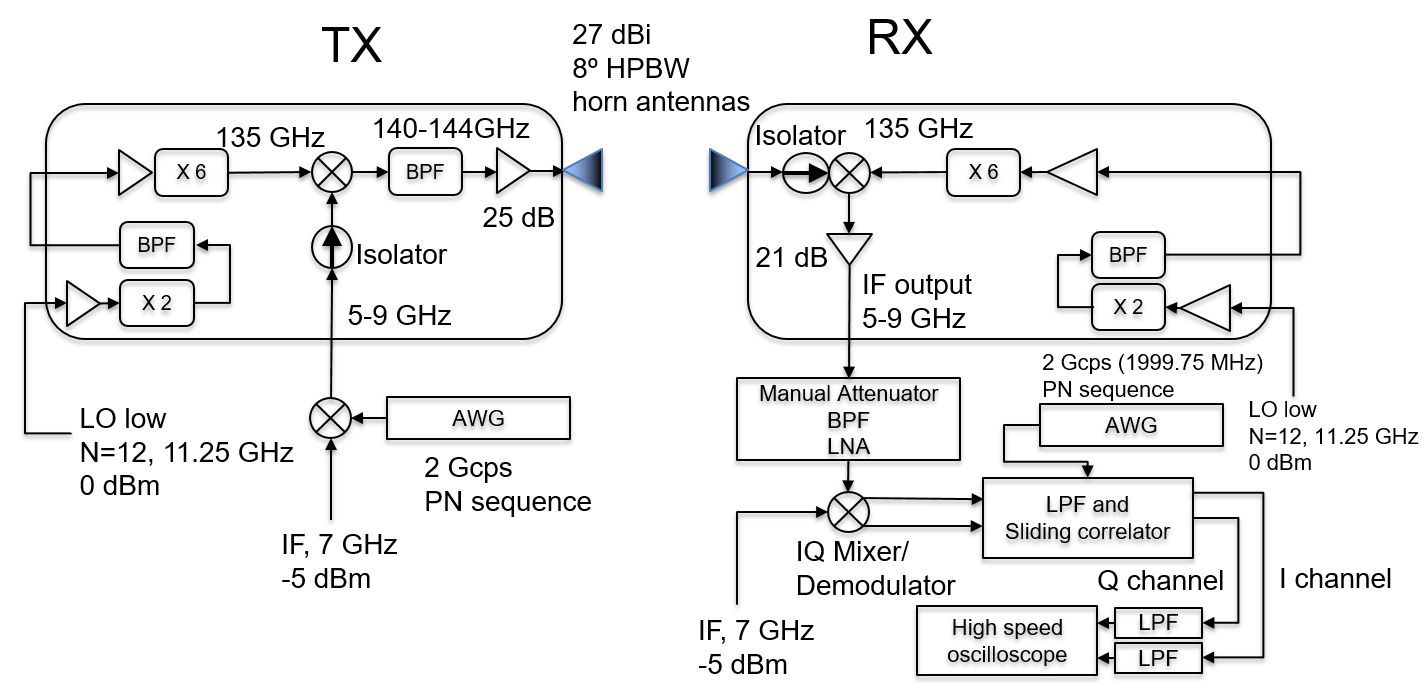}
	\vspace{-1.5 em}
	\caption{\textcolor{black}{Block diagram of 140 GHz broadband channel sounder system at NYU WIRELESS.}}
	\vspace{-1.5 em}
	\label{fig:block}
\end{figure}

Another previous work on D-band propagation measurements and characterization were conducted at Georgia Institute of Technology in a very close-in environment around a personal computer \cite{kim2015Dband,cheng2017comparison}. The D-band channel sounder was based on a VNA, where the LO frequency ranged from 11 to 17 GHz was upconverted to the D-band 110-170 GHz. With a transmit power of 0 dBm and 23 dBi gain pyramidal horn antennas at both the TX and RX, the channel sounder system had a dynamic range of 90 dB for a 100 kHz intermediate frequency (IF) bandwidth \cite{kim2015Dband}. The measurement data was collected over TX-RX (TR) separation distances $d$ that varied from 35.56 to 86.36 cm due to the limited dynamic range of the system \cite{cheng2017comparison}.

Indoor directional path losses at 30 GHz, 140 GHz, and 300 GHz were compared using four different path loss models in \cite{cheng2017comparison}, i.e. the single-frequency FI model, the single-frequency CI free space reference distance model (with a reference distance of 0.1 m), the multi-frequency CI model with a frequency-dependent term (CIF), and the multi-frequency alpha-beta-gamma (ABG) model \cite{APS2,maccartney2015indoor,3GPP2017}. It was shown that although all of these four LOS path loss models offered PLEs close to 2.0, the multi-frequency CIF and ABG model had better stability for PLE and standard deviation than the single-frequency CI and FI models \cite{cheng2017comparison}. However, due to the small dynamic range of the channel sounder system, path loss measurement distances were limited to 1.8 m.

\begin{figure}    
	\centering
	\includegraphics[width=0.5\textwidth]{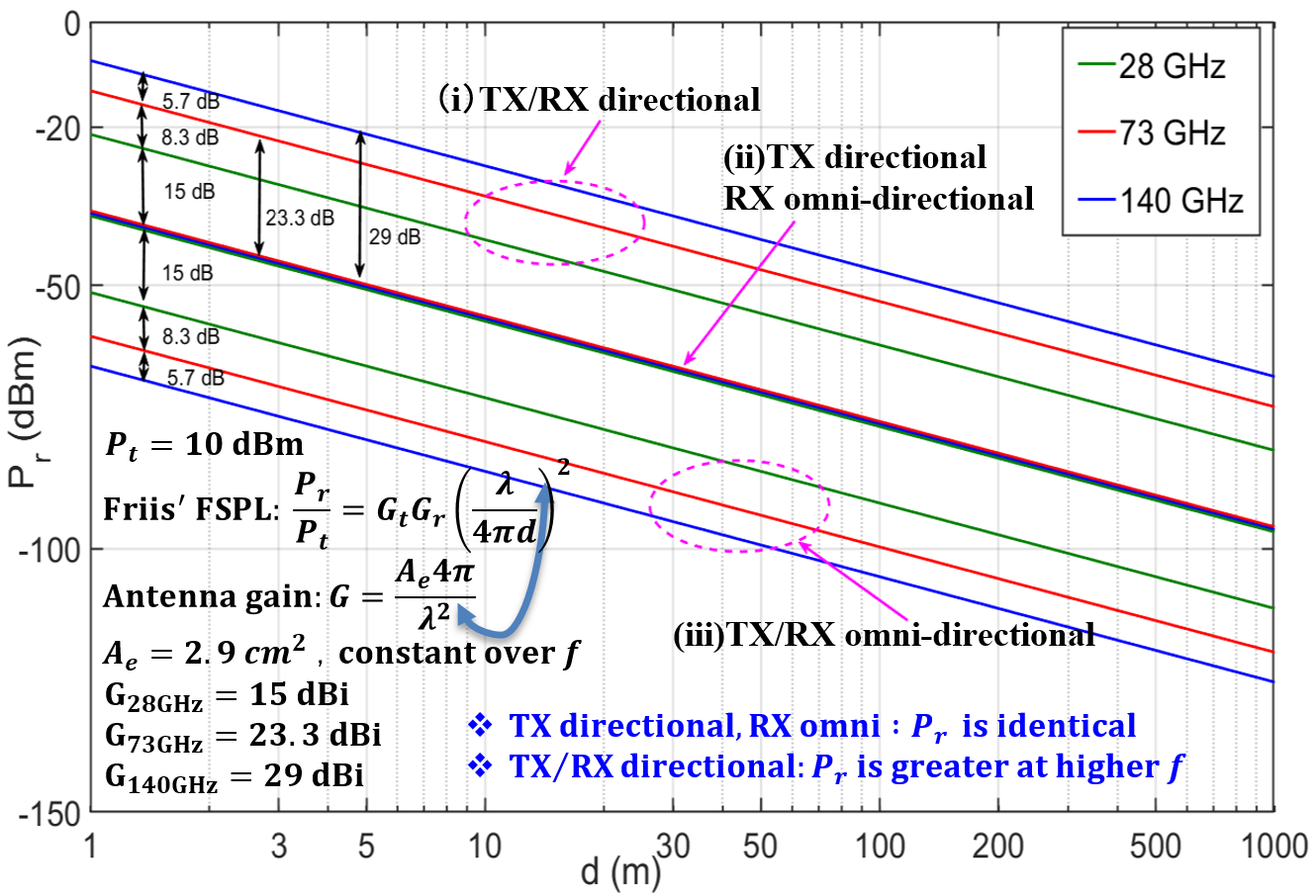}
	\vspace{-1.5 em}
	\caption{Received power vs. distance with (i) TX and RX are both directional, (ii) TX is directional but RX is omni-directional, and (iii) both TX and RX are omni-directional at 28 GHz, 73 GHz, and 140 GHz. Directional antennas with equal effective aperture ($A_e = $ 2.9 $\text{cm}^2$) at both TX and RX have much less path loss at higher frequencies (see Ch.3 in \cite{Rap15a}).}
	\vspace{-1.5 em}
	\label{fig:FSPLPr}
\end{figure}

\section{A novel 140 GHz channel sounder system}

The 140 GHz channel sounding system at NYU WIRELESS can support both a wideband sliding correlator mode and a real-time spread spectrum mode, which are suitable for both long-distance propagation measurements with angular/delay spread  and short-range dynamic channel measurements for Doppler and rapidly fading characterization, respectively \cite{Mac17JSACb}. The block diagram of the dual conversion 140 GHz sliding correlator system is shown in Fig. \ref{fig:block}, where a 2 Giga-chip-per-second (Gcps) pseudorandom noise (PN) sequence of 2047 chips in length (11 bits) is generated by an arbitrary waveform generator (AWG, Tektronix AWG70002A). The PN sequence is modulated by an IF signal centered at 7 GHz, and then the 4 GHz wide null-to-null bandwidth signal at IF enters the RF upconverter. The LO signal is set at 11.25 GHz and passes through a band pass filter (BPF) and a $\times$ 12 frequency multiplication. Then, the LO and IF signals are mixed to generate the output RF signal centered at 142 GHz ($11.25\times 12+7=142$ GHz) and the image frequencies are filtered off by a 140 GHz band BPF (139-145 GHz, the 3 dB passband bandwidth). After being amplified by a 25 dB gain power amplifier,  the RF signal is radiated through a 27 dBi gain rotatable horn antenna with a 8\textdegree~HPBW in both azimuth and elevation planes.

At the RX, the RF signal is captured by the rotatable horn antenna and then down converted by the 135 GHz LO signal. \textcolor{black}{The down converted signal passes through a variable attenuator, a BPF, a low noise amplifier (LNA), and is I/Q demodulated with the 7 GHz IF signal to baseband. The demodulated I and Q signals are correlated with an identical PN sequence but at a slightly offset rate, providing autocorrelation processing gain at the expense of a slightly longer acquisition time (on the order of tens of ms) \cite{Mac17JSACb, Rap02a, mac17ICC}. With the processing gain, a sliding correlator based channel sounder has a much larger dynamic range than VNA methods \cite{Rap02a, Mac17JSACb}. For the 140 GHz sliding correlator, the 11-bit PN sequence provides a 66 dB autocorrelation processing gain ($20 \times \log_{10}2^{11}$) and the maximum measurable dynamic range is 145 dB (verified by measurements) with a transmit RF output power of 0 dBm using 27 dBi horn antennas at both TX and RX. Theoretically, the path loss in free space decreases quadratically as frequency increases, so long as the physical size of the antenna (effective aperture) is kept constant over frequency at both link ends \cite{APS2, Rap02a}. The astounding result of improved coverage at higher frequencies is clear in Fig. \ref{fig:FSPLPr}, where the received power at 140 GHz free space is 5.7 dB greater than at 73 GHz and 14 dB greater than at 28 GHz for the same TX output power and for identical physical antenna areas at all frequencies (see Ch.3 in \cite{Rap15a}). }

The measurable path loss range of the 140 GHz system is less than that used in \cite{Mac17JSACb, mac17ICC} because of the smaller output power provided by the power amplifier (0 dBm at 140 GHz as compared to 30 dBm at 28 GHz \cite{rappaport2013millimeter}), due to the present lack of amplifier technology at such high frequency). Two separate high stability Rubidium (Rb) clocks are used at the TX and RX for synchronization without requiring any cables between the TX and RX.  The TX and RX antennas are both mounted on gimbals which can be swept 360\textdegree~in azimuth plane and 120\textdegree~in elevation plane in 1\textdegree/step, so that dual-directional information (AOA and AOD) can be obtained for MIMO and directionality analysis.

\section{Free space path loss and indoor penetration measurements at 140 GHz.}

\begin{figure}    
	\centering
	\includegraphics[width=0.5\textwidth]{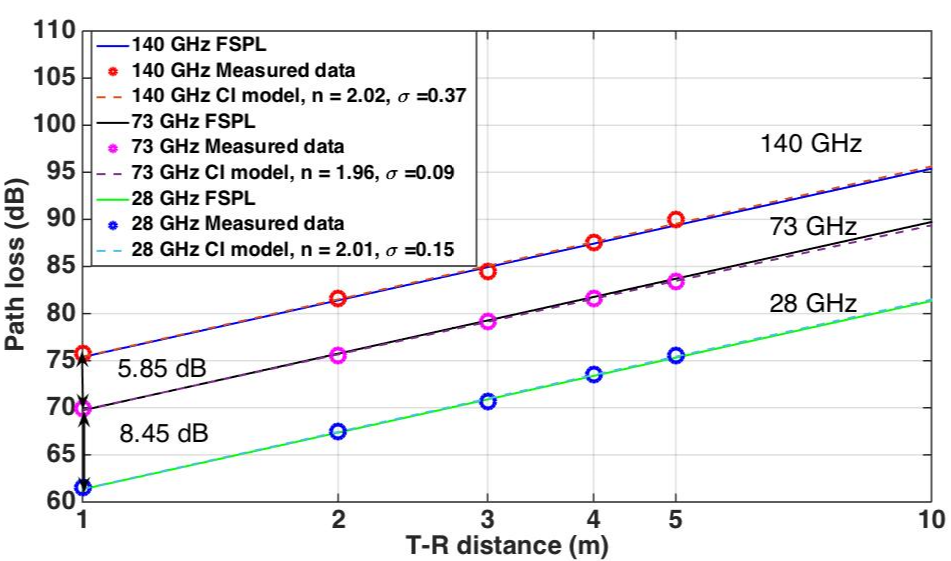}
		\vspace{-1.5 em}
	\caption{28, 73 and 140 GHz free space path loss (after subtracting out all antenna gains) verification measurements at distances of 1, 2, 3, 4, and 5 m.}
		\vspace{-1.5 em}
	\label{fig:140fspl}
\end{figure}

140 GHz free space path loss verification measurements were conducted at T-R separation distances of 1, 2, 3, 4, and 5 m using the recently proposed standard calibration and verification method taught in \cite{xing18VTC}, and the results are shown after removal of antenna gains in Fig. \ref{fig:140fspl} together with 28 GHz and 73 GHz measurement data. The measured path loss at 140 GHz agrees with Friis FSPL equation \cite{friis1946note}, and the CI path loss model with 1 m reference distance \cite{rappaport2013millimeter} agrees well with the measured data and FSPL model, which shows the CI path loss model with 1 m reference distance still holds for 140 GHz (above 100 GHz).


As shown in Fig. \ref{fig:140fspl}, the measured path loss difference between 73 GHz and 140 GHz at the same distance (e.g., 1 m) is 5.85 dB which is extremely close to the theoretical value calculated by Friis FSPL equation ($20\times10\log_{10}\frac{140}{73}=5.66 $ dB), indicating high accuracy of the channel sounder system. The measured path loss difference is 8.45 dB between the 28 and 73 GHz measurements at the same distance, which is virtually identical to Friis FSPL equation ($20\times10\log_{10}\frac{73}{28}=8.32$ dB) \cite{xing18VTC}.

Indoor and outdoor environments at D-band and THz frequencies need to be extensively investigated for the impact of penetration loss of common materials, as knowledge of such loss shall be essential to predict indoor and outdoor-to-indoor path loss needed for the design and installation of future 5G mmWave wireless systems in and around buildings \cite{jacque2016indoor,maccartney2015indoor}. Penetration measurements at 140 GHz were conducted at the NYU WIRELESS research center, where T-R separation distances of 3, 4, and 5 m were used and the TX/RX antenna heights were 1.5 m (see Fig. 3 in \cite{xing18VTC}). The penetration losses were calculated as:

\begin{equation}\label{equ:LVV}
\begin{split}
L_{V-V}\text{[dB]} = P_t\text{[dBm]}-P_{r-V}^{MUT}(d)\text{[dBm]} + G_{TX}\text{[dBi]}\\
+G_{RX}\text{[dBi]}-PL_{V-V}(d)\text{[dB]},
\end{split}
\end{equation}
where $P_{r-V}^{MUT}(d)$  is the co-polarized received powers in dBm at distance $d$ in meters at the output of the RX antenna with the material under test (MUT) between the TX and RX antenna, $L_{V-V}\text{[dB]}$ is the co-polarized material penetration loss, $P_t\text{[dBm]}$ is the transmitted power into the TX antenna, $G_{TX}\text{[dBi]}$ and $G_{RX}\text{[dBi]}$ are the TX and RX antenna gains, respectively, and $PL_{V-V}(d)\text{[dB]}$ is the free space path loss at distance $d$.

\begin{figure}    
	\centering
	\includegraphics[width=0.45\textwidth]{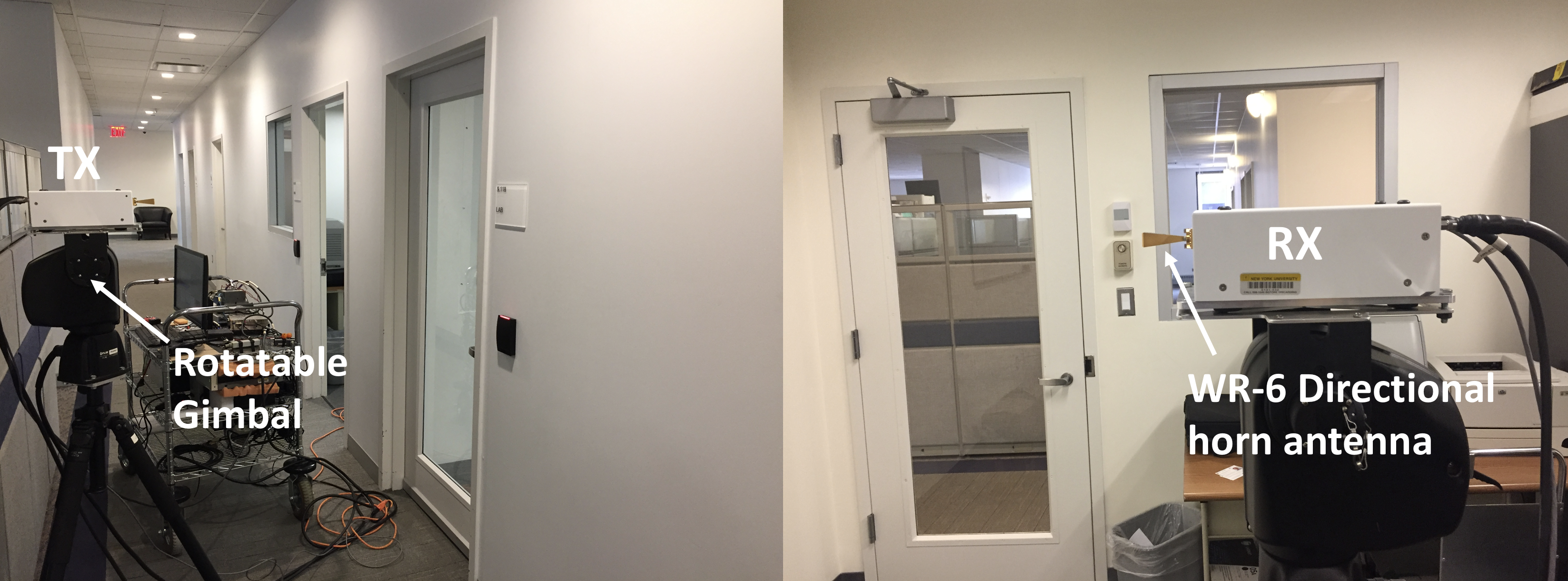}

	\caption{Penetration loss measurements at 140 GHz with directional horn antennas.}

	\label{fig:140CS}
\end{figure}

\begin{table*}[!ht]
	\renewcommand{\arraystretch}{1.1}
	\centering
	\caption{Comparison of drywall and clear glass penetration loss at 28, 73, and 140 GHz \cite{jacque2016indoor,zhao2013icc}} \label{tab:PE}
	\newcommand{\tabincell}[2]{\begin{tabular}{@{}#1@{}}#2\end{tabular}}
	\begin{tabular}{|c|c|c|c|c|}
		\hline
		\tabincell{c}{\textbf{Frequency} \textbf{(GHz)} }&\tabincell{c}{\textbf{MUT}}&\tabincell{c}{\textbf{Thickness} \textbf{(cm)} }&\tabincell{c}{\textbf{Penetration Loss} \textbf{ (dB)} }&\tabincell{c}{\textbf{Avg. Penetration Loss} \textbf{ (dB/cm)} }\\
		\hline
		\tabincell{c}{28 \cite{rappaport2013millimeter,zhao2013icc}}&\tabincell{c}{Clear glass A}&\tabincell{c}{ 1.2}&\tabincell{c}{3.6}&\tabincell{c}{3.0}\\
		\hline
		\tabincell{c}{28 \cite{rappaport2013millimeter,zhao2013icc}}&\tabincell{c}{Clear glass B}&\tabincell{c}{ 1.2}&\tabincell{c}{3.9}&\tabincell{c}{3.25}\\
		\hline
		\tabincell{c}{28 \cite{rappaport2013millimeter,zhao2013icc}}&\tabincell{c}{Drywall A}&\tabincell{c}{38.1}&\tabincell{c}{6.8}&\tabincell{c}{0.18}\\
		\hline
		\hline
		\tabincell{c}{ 73 \cite{jacque2016indoor}}&\tabincell{c}{Clear glass C}&\tabincell{c}{0.6}&\tabincell{c}{7.72 }&\tabincell{c}{12.87}\\
		\hline
		\tabincell{c}{ 73 \cite{jacque2016indoor}}&\tabincell{c}{Clear glass D}&\tabincell{c}{0.6}&\tabincell{c}{7.1 }&\tabincell{c}{11.83}\\
		\hline
		\tabincell{c}{ 73 \cite{jacque2016indoor}}&\tabincell{c}{Drywall B}&\tabincell{c}{14.5}&\tabincell{c}{10.06 }&\tabincell{c}{0.73}\\
		\hline
		\hline
		\tabincell{c}{ 140 }&\tabincell{c}{Clear glass C}&\tabincell{c}{0.6}&\tabincell{c}{ 8.24 }&\tabincell{c}{13.73}\\
		\hline
		\tabincell{c}{ 140 }&\tabincell{c}{Clear glass D}&\tabincell{c}{0.6}&\tabincell{c}{ 9.07 }&\tabincell{c}{15.12}\\
		\hline
		\tabincell{c}{ 140 }&\tabincell{c}{Glass door (Front door)}&\tabincell{c}{1.3}&\tabincell{c}{ 16.2 }&\tabincell{c}{12.46}\\
		\hline
		\tabincell{c}{ 140 }&\tabincell{c}{Drywall B}&\tabincell{c}{14.5}&\tabincell{c}{ 15.02 }&\tabincell{c}{1.04}\\
		\hline
		\tabincell{c}{ 140 }&\tabincell{c}{Drywall with Whiteboard}&\tabincell{c}{17.1}&\tabincell{c}{ 16.69 }&\tabincell{c}{0.98}\\
		\hline		
	\end{tabular}
\end{table*}


Common materials like drywall, clear glass, and a glass door were measured at 140 GHz. To explore variations of different samples, clear glass with thicknesses of 0.6 cm and 1.2 cm at different locations were chosen to be the MUT, and the drywall with different thicknesses of 14.5 cm, 17.1 cm, and 38.1 cm was chosen to be measured. Fig. \ref{fig:140CS} shows an example of the penetration loss measurements with the 140 GHz channel sounder for different common materials, with results given in Table \ref{tab:PE}.

Previous work at 28 and 73 GHz \cite{jacque2016indoor, zhao2013icc} showed that a 1.2-cm-thick clear glass penetration loss was 3.6 dB (3.9 dB at another location with the same thickness) at 28 GHz \cite{zhao2013icc} and a 0.6-cm-thick clear glass would introduce 7.72 dB penetration loss (7.1 dB at another location with the same thickness) at 73 GHz \cite{jacque2016indoor}. However, the penetration loss of clear glass with a thickness of 0.6 cm at 140 GHz was 9.07 dB (8.24 dB at another location with the same thickness), see Table \ref{tab:PE} for comparisons. Average penetration loss (dB/cm) was calculated by dividing the penetration loss by the MUT thickness, which shows the penetration property of each material.

 Measurements in \cite{jacque2016indoor,zhao2013icc,rappaport2013millimeter} presents the average penetration loss of clear glass is 3.2 dB/cm at 28 GHz, while it is 12.3 dB/cm and 14 dB/cm at 73 and 140 GHz, respectively, showing that the penetration loss increases with frequencies. A similar result holds for the drywall, where the average penetration loss is 0.18 dB/cm at 28 GHz, but increases to 0.73 dB/cm at 73 GHz and is 1.04 dB/cm at 140 GHz. The same material at different locations has similar average penetration loss, which confirms the consistancy of the measurements. As the penetration loss at higher frequencies is greater, the penetration loss of clothing and garment materials could be expected to be non-negligible at THz frequencies, even though the mmWave attenuation of most garment materials is shown to be negligible in \cite{wu2015safe,wu15ICC}.
 

\section{Planned indoor propagation measurements}

Indoor 140 GHz broadband wireless propagation measurements are planned to be conducted in a multipath-rich indoor environment on the $9^{th}$ floor of 2 MetroTech Center, which is a typical indoor environment including a hallway, meeting rooms, cubical office, laboratory and open areas. The purpose of the 140 GHz indoor measurement campaign is to collect data for various locations and antenna polarizations for creation of broadband statistical channel model that is frequency dependent and can be implemented in a format similar to NYUSIM \cite{sun2017a}, where the model shall be formed from various field propagation measurements and existing 28 and 73 GHz indoor data \cite{maccartney2015indoor}. The 140 GHz indoor measurement campaign will include the same measurement locations as used at 28 and 73 GHz \cite{maccartney2015indoor}, providing 48 TX-RX combinations ranging from 3.9 to 45.9 m. The processed data and resulting models will help with the design and investigation of mmWave indoor wireless access networks, future gigabyte WiFi, and Internet of things. 

Position localization in wideband mmWave systems is an interesting topic and a future use case of mmWave and THz wireless networks \cite{ojas2018globecomm}. The 140 GHz measurement campaign (with 4 GHz RF bandwidth) together with the previous 28 and 73 GHz measurements can support indoor and outdoor channel modeling and the study of precise localization algorithms. \textcolor{black}{The designed measurements will also be used for spatial consistency analysis and implementation, to ensure that the resulting indoor channel models experience smooth channel transitions when moving in a local area \cite{ju18VTC, ju18GC}.}

\section{Conclusion}
This paper summarized wireless communication research and rulemakings above 100 GHz, overviewed the existing propagation measurements at D-band (110-170 GHz), provided the architecture of NYU WIRELESS 140 GHz channel sounder system, and presented preliminary penetration loss measurements at 140 GHz for various building materials. Penetration loss and average penetration loss of common materials at 140 GHz, which are not well investigated, are measured and compared with those at 28 and 73 GHz. A planned 140 GHz indoor measurement campaign is proposed and, together with the previous 28 and 73 GHz indoor measurements conducted at NYU WIRELESS, they will be used to form statistical indoor channel models for various TX and RX antenna configurations and polarizations at multiple frequencies. The processed data and resulting models will help with mmWave indoor wireless network design, position localization studies, and future gigabyte WiFi with Internet of things. 

\section{Acknowledgments}
This research is supported by the NYU WIRELESS Industrial Affiliates Program and two National Science Foundation (NSF) Research Grants: 1702967 and 1731290.

\bibliographystyle{IEEEtran}
\bibliography{globecom140GHz}

\end{document}